\documentclass[graybox]{svmult}

\usepackage{mathptmx}       
\usepackage{helvet}         
\usepackage{courier}        
\usepackage{type1cm}        
\usepackage{makeidx}         
\usepackage{graphicx}        
\usepackage{multicol}        
\usepackage[bottom]{footmisc}

\makeindex             

\begin{document}

\title*{Molecular and atomic gas in the young TeV $\gamma$-ray SNRs RX J1713.7$-$3946 and RX J0852.0$-$4622; evidence for the hadronic production of $\gamma$-rays}
\titlerunning{Molecular and atomic gas in the young TeV $\gamma$-ray SNRs}
\author{Yasuo Fukui}
\institute{Yasuo Fukui,\at Department of Physics, Nagoya University, Furo-cho, Chikusa-ku, Nagoya, 464-8601, Japan,\at \email{fukui@a.phys.nagoya-u.ac.jp}}
%
%
\maketitle

\abstract*{The interstellar molecular clouds are the site of star formation and also the target for the cosmic ray protons to produce $\gamma$-rays via the hadronic process. The interstellar atomic gas is enveloping the molecular clouds and may also be dense enough to affect the $\gamma$-ray production. In this Chapter, some of the basic properties of the interstellar gas both in molecular and atomic forms will be reviewed. Then, it is presented that two young TeV $\gamma$-ray SNRs, RX J1713.7$-$3946 and RX J0852.0$-$4622, show good spatial correspondence between the $\gamma$-rays and the interstellar protons. The good spatial correspondence provides a support for the hadronic origin of the $\gamma$-rays in these SNRs. It is emphasized that both molecular and atomic hydrogen plays a role as targets for cosmic ray (CR) protons. The clumpy distribution of the target interstellar medium (ISM) protons is crucial in the interaction of the supernova shocks with the ISM, whereas models with uniform ISM distribution are not viable. Finally, it is suggested that the dense atomic gas without molecules may occupy the dominant part of the dark gas in the local ISM.}

\abstract{The interstellar molecular clouds are the site of star formation and also the target for the cosmic ray protons to produce $\gamma$-rays via the hadronic process. The interstellar atomic gas is enveloping the molecular clouds and may also be dense enough to affect the $\gamma$-ray production. In this Chapter, some of the basic properties of the interstellar gas both in molecular and atomic forms will be reviewed. Then, it is presented that two young TeV $\gamma$-ray SNRs, RX J1713.7$-$3946 and RX J0852.0$-$4622, show good spatial correspondence between the $\gamma$-rays and the interstellar protons. The good spatial correspondence provides a support for the hadronic origin of the $\gamma$-rays in these SNRs. It is emphasized that both molecular and atomic hydrogen plays a role as targets for cosmic ray (CR) protons. The clumpy distribution of the target interstellar medium (ISM) protons is crucial in the interaction of the supernova shocks with the ISM, whereas models with uniform ISM distribution are not viable. Finally, it is suggested that the dense atomic gas without molecules may occupy the dominant part of the dark gas in the local ISM.}

\section{Introduction}
\label{sec:1}
Recent progress in $\gamma$-ray imaging of supernova remnants (SNRs) is remarkable. In particular, several $\gamma$-ray telescopes, H.E.S.S., VERITAS, MAGIC, Fermi and AGILE, are obtaining high quality $\gamma$-ray images of high-energy objects including SNRs and PWNs in the Galaxy in the last decade (e.g., Gast et al. 2012; Hui 2010; Zanin 2009; Castro $\&$ Slane 2010; Giuliani et al. 2011). It was originally suggested in the middle of the 20th century that we are able to observe the $\gamma$-rays produced by the hadronic interaction between the cosmic ray (CR) protons and the interstellar protons which produce neutral pions decaying into $\gamma$-rays (Hayakawa 1956). 

The history of observations of the interstellar medium (ISM) spans over more than a half century and observations of the ISM have been made mainly at radio, infrared and optical wavelengths. The H{\sc i} observations at 21 cm and the CO observations at 2.6 mm provide powerful tools to probe the major neutral component of the ISM. In particular, early CO surveys allowed us to make a comparison between the $\gamma$-rays and the molecular protons (Dame et al. 2001). The COS-B satellite obtained the $\gamma$-ray distribution of the whole sky in the 100 MeV--10 GeV range (Mayer et al. 1980). The correlation between the $\gamma$-rays and CO is good at a degree-scale resolution and was used to derive a $X$ factor (=$N({\mathrm{H_2}}$)/$W(^{12}$CO)) that relates the CO intensity to the proton density (Lebrun et al. 1983; Bloemen et al. 1986; Strong et al. 1988).

The TeV $\gamma$-ray astronomy with ground-based atmospheric Cerenkov telescopes including H.E.S.S. opened a new era by detecting the highest energy $\gamma$-rays in the Galaxy. The TeV $\gamma$-rays may be produced by the CR protons whose energy is close to the knee and have a potential to probe the highest energy Galactic CRs and the acceleration site. The H.E.S.S. survey of the Galactic plane which detected more than 50 $\gamma$-ray sources was compared extensively with the NANTEN CO observations and the region of the W28 SNR was found to show the best spatial correspondence with the TeV $\gamma$-rays as shown in Figure 1 (Aharonian et al. 2008). It however remained as a puzzle why the other $\gamma$-ray sources do not show clear correspondence with the CO. For instance, the most remarkable TeV $\gamma$-ray SNRs, RX J1713.7$-$3946 and RX J0852.0$-$4622, were not identified conclusively to be associated with CO (e.g., Aharonian et al. 2007b). In this Chapter, we show that this puzzle has been solved by a careful analysis of the ISM protons including atomic protons.

\begin{figure*}[t]
\begin{center}
\includegraphics[width=117mm,clip]{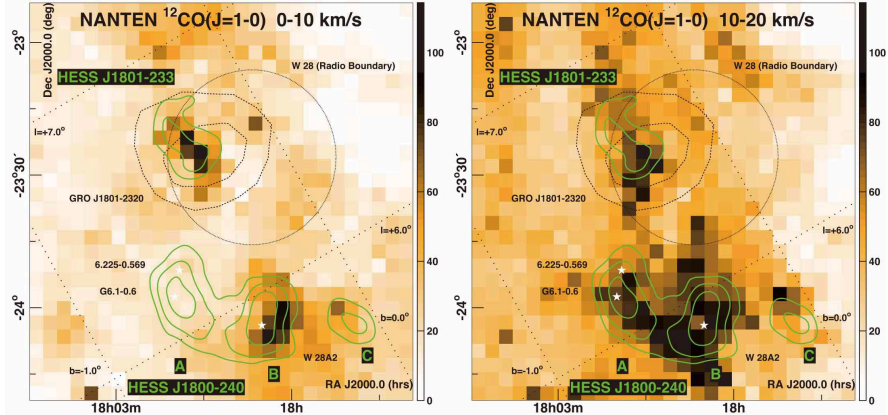}
\caption{NANTEN $^{12}$CO($J$=1--0) integrated intensity distribution toward the W28 region (linear scale in K km s$^{-1}$) with TeV $\gamma$-ray significance contours overlaid ($green$) 4, 5, 6$\sigma$ levels (Aharonian et al. 2008 Figure 2). The velocity ranges are 0--10 km s$^{-1}$ ($left$ $panel$) and 10--20 km s$^{-1}$ ($right$ $panel$). The radio boundary of W28, GRO J1801$-$2320 and the location of the H{\sc ii} region W28A2 ($white$ $stars$) are shown in the in the panels.}
\end{center}
\end{figure*}%

\section{Interstellar medium}
\label{sec:2}
The interstellar medium consists of gas and dust. The mass ratio of the gas and dust is 100:1. The gas is either neutral or ionized. The ionized gas is a minor component in mass, since recombination is rapid in the Galactic disk for the general ionizing UV radiation field from OB stars. Therefore, the neutral gas is the dominant component in mass. The neutral gas is mainly either atomic or molecular hydrogen. Because of the higher cooling rate at higher densities, the denser gas tends to be cooler; the cooling rate is proportional to density squared, whereas the heating rate linearly depends on density.

The atomic gas (H{\sc i} gas) has a density from 0.1 cm$^{-3}$ to 100 cm$^{-3}$ with kinetic temperature from 30 K to 3000 K. It is generally believed that such interstellar medium consists of two dynamically stable phases, either ``high temperature and low density'' and ``low temperature and high density'' (Spitzer 1978). It is yet possible that a significant amount of transient H{\sc i} gas at an intermediate density range exits, since the ISM gas is highly turbulent and far from the dynamically equilibrium (e.g., Koyama $\&$ Inutsuka 2002). It is generally assumed that the H{\sc i} emission is optically thin as suggested by the non-flat H{\sc i} line profiles (Dickey $\&$ Lockman 1990), whereas it is possible that there exits opaque H{\sc i} emission from dense H{\sc i}. Such H{\sc i} gas may be important in a shell compressed by the stellar winds in SNRs, although the nature of such H{\sc i} gas has not been discussed often in the literature.

The molecular gas has density greater than 10$^3$ cm$^{-3}$ with typical temperature of 10 K (e.g., Goldsmith 1987). The low temperature is in part due to the efficient molecular cooling and the strong shielding of stellar photons by dust grains. The major heating of the molecular clouds is made by CR protons of a 100 MeV range. The main constituent of the molecular clouds is H$_2$ but H$_2$ does not have a radiative transition observable at low temperature of 10 K. We use usually the rotational transition of CO at 2.6 mm wavelength to observe the molecular clouds. The CO transition (rotational quantum number $J$=1--0) can be excited via collision with H$_2$ at 10 K. H$_2$ is formed from H{\sc i} by reactions on dust surface and is then liberated into the interstellar space. There is no reliable tracer for a density range, approximately 100--1000 cm$^{-3}$. In this range, CO/H$_2$ may not be formed yet and H{\sc i} may be optically thick. It is therefore difficult to see the phase transition from H{\sc i} to H$_2$, while future studies may be able to establish CI or CII transitions to trace such a density regime.

The molecular clouds are clumpier than the H{\sc i} gas and the mass of a molecular cloud ranges from 100 $M_{\odot}$ to 10$^7$ $M_{\odot}$. The large clouds having mass greater than 10$^5$ $M_{\odot}$ are called giant molecular clouds (GMCs). The total number of GMCs in the Galaxy is estimated to be 3000. GMCs are nearly self-gravitating (e.g., Fukui and Kawamura 2010). They have long lifetime of a few 10 Myrs (Kawamura et al. 2009) and are the formation site of stars of various mass. Ultraviolet radiation and stellar winds of high-mass stars formed in the GMCs ionize and eventually disperse GMCs.  
Such high-mass stars are the progenitors of SN explosions where the CR rays are accelerated.

\section{Gamma-ray SNRs (supernova remnants)}
\label{sec:3}
The interstellar medium (ISM) is crucial in producing high-energy radiation from supernova remnants. In particular, the protons in the ISM play a role as targets for CR protons to produce $\gamma$-rays. CR protons do not emit electromagnetic radiation, whereas a reaction between CR protons and ISM protons produces a neutral pion which decays into two $\gamma$-rays. This is the so called hadronic process of $\gamma$-ray production and provides a unique probe for the CR protons. Another reaction that produces $\gamma$-rays is the leptonic process in which CR electrons energize low energy photons into $\gamma$-rays via the inverse Compton effect. Such CR electrons are observed directly via non-thermal synchrotron emission. One of the most important issues related to the $\gamma$-ray SNRs is if the $\gamma$-rays are produced by the hadronc process or the leptonic process. It is of vital importance to establish the hadronic-origin $\gamma$-rays in order to verify the origin of the CRs, a long standing puzzle since 1912 when the CRs were discovered by V. Hess.

If the hadronic process is working, we expect that the spatial distribution of the $\gamma$-rays corresponds to that of the ISM protons. Some of the previous efforts however did not find reasonable correspondence between the $\gamma$-rays and the ISM in the two most outstanding TeV $\gamma$-ray SNRs, RX J1713.7$-$3946 and RX J0852.0$-$4622 (Aharonina et al. 2006b, 2007b). These previous works used only mm-wave CO data to estimate the ISM proton distribution. It may however be necessary to include atomic protons also as the target protons if their density is high enough (Section 2). We employ the TeV $\gamma$-ray observations by H.E.S.S. at angular resolution of 0.1 degrees, as well as the CO and H{\sc i} data obtained with the NANTEN2 4m telescope and ATCA (the Australia Telescope Compact Array) combined with the Parkes 64 m telescope, respectively. We present the results of comparisons between the TeV $\gamma$-rays and the ISM protons in the two SNRs, RX J1713.7$-$3946 and RX J0852.0$-$4622, to demonstrate that the TeV $\gamma$-ray distribution well corresponds to the ISM proton distribution where both atomic and molecular protons are taken into account (Fukui et al. 2012, 2013).

\subsection{RX J1713.7$-$3946}
\label{subsec:3-1}
RX J1713.7$-$3946 is the brightest TeV $\gamma$-ray SNR detected in the Galactic plane survey with H.E.S.S. (Aharonian et al. 2006a)  and is a most promising candidate where the origin of the $\gamma$-rays may be established. The $\gamma$-rays higher than 10 TeV is detected here, suggesting that the CR protons responsible may have energy close to the knee, 10$^{15}$ eV, if the $\gamma$-rays are hadronic. The SNR was discovered in X-rays with ROSAT (Pfeffermann $\&$ Aschenbach 1996) and soon the X-rays are found to be non-thermal synchrotron emission with no thermal features (Koyama et al. 1997). TeV $\gamma$-rays were first detected by CANGAROO (Enomoto et al. 2002) and H.E.S.S. resolved the shell-like TeV $\gamma$-ray distribution (Aharonian et al. 2004, 2006b and 2007a). Considerable work has been devoted to explore the $\gamma$-ray emission mechanisms (Aharonian et al. 2006b; Porter et al. 2006; Katz $\&$ Waxman 2008; Berezhko $\&$ V$\mathrm{\ddot{o}}$lk 2008; Ellison $\&$ Vladimirov 2008; Tanaka et al. 2008; Morlino et al. 2009; Acero et al. 2009; Ellison et al. 2010; Patnaude et al. 2010; Zirakashvili $\&$ Aharonian 2010; Abdo et al. 2011; Fang et al. 2011).

\begin{figure*}[b]
\begin{center}
\includegraphics[width=78mm,clip]{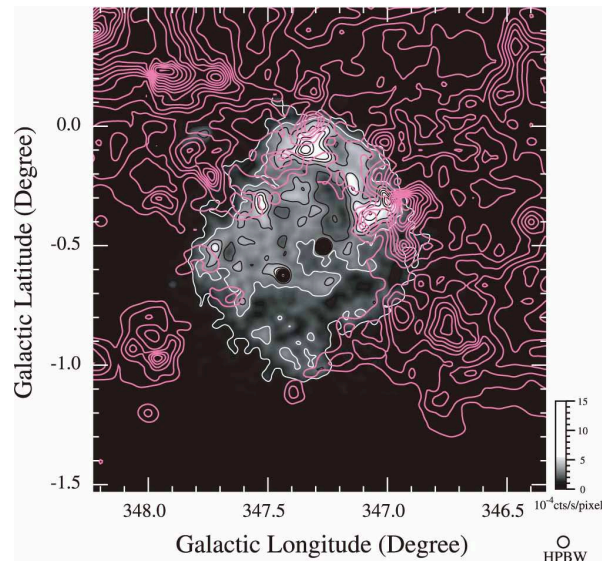}
\caption{Overlay map in Galactic coordinates showing a supernova remnant (SNR), RX J1713.7$-$3946, in gray scale [ROSAT PSPC X-ray Survey (Slane et al. 1999); from ROSAT archive database] and the intensity distribution of CO($J$=1--0) emission in $magenta$ $contours$. The intensity is derived by integrating the CO spectra from $-$11 to $-$3 km s$^{-1}$ , which is considered to be a velocity component interacting with the SNR. The $lowest$ $contour$ level and interval of CO are 4 K km s$^{-1}$ (Fukui et al. 2003).}
\end{center}
\end{figure*}%

The molecular gas interacting with the SNR was discovered in the CO $J$=1--0 emission in $V_{\mathrm{LSR}}$, the velocity with respect to the local standard of rest, around $-7$ km s$^{-1}$. Figure 2 shows that the distribution of the CO emission in RX J1713.7$-$3946 well correlates with the X-ray distribution (Fukui et al. 2003); the northwestern rim of the X-ray coincides with the most prominent CO peaks (Fukui et al. 2003, Moriguchi et al. 2005, Fukui 2008, and Sano et al. 2010). This X-ray distribution is now interpreted as caused by the interaction between the shock front and the molecular clouds as modeled in the magneto-hydrodynamical (MHD) numerical simulations by Inoue et al. (2012). The correlation provides a robust verification of the physical association of the CO clouds with the SNR shell.

The distance of the SNR was determined to be 1 kpc by using the flat rotation curve of the Galaxy (Fukui et al. 2003). Studies of X-ray absorption suggested a similar distance 1 kpc under an assumption of uniform foreground gas distribution (Koyama et al. 1997), but the local bubble of H{\sc i} located by chance toward the SNR makes the X-ray absorption uncertain in estimating the distance (Slane et al.1999). A subsequent careful analysis of the X-ray absorption also favors the smaller distance (Cassam-Chenai et al. 2004). At 1 kpc the SNR has a radius of 9 pc and an age of 1600 yrs (Fukui et al. 2003; Wang et al. 1997) and the expanding shock front has a speed of 3000 km s$^{-1}$ (Zirakashvili $\&$ Aharonian 2010; Uchiyama et al. 2003, 2007). Most recently, Sano et al. (2010) showed that the SNR harbors the star forming dense clump peak C (Figure 3b) and argued that the X-ray intensity is significantly enhanced around the clump due to the shock interaction, reinforcing the association of the molecular gas.

The molecular gas associated with the SNR opened a unique possibility to identify target protons in the hadronic process. We expect the $\gamma$-ray distribution mimics that of the interstellar target protons if the hadronic process is working and if the CR distribution is uniform within the SNR. A detailed comparison between the ISM protons and the high-resolution $\gamma$-ray image of H.E.S.S. is therefore a useful test of the hadronic scenario.

\begin{figure*}[t]
\begin{center}
\includegraphics[width=117mm,clip]{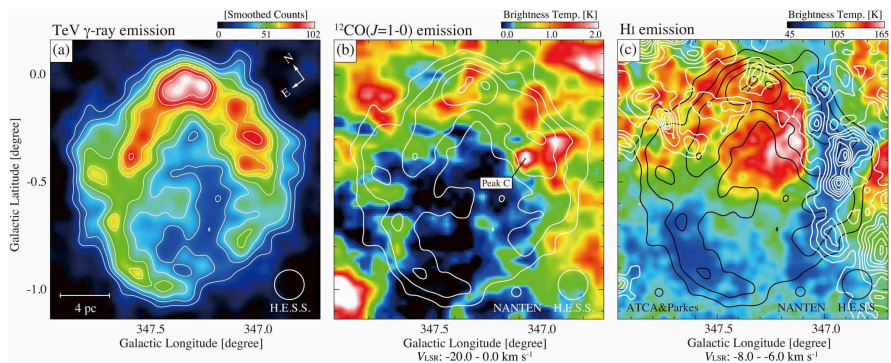}
\caption{(a) The H.E.S.S. TeV $\gamma$-ray distribution of RX J1713.7$-$3946 in smoothed excess counts above the cosmic-ray background (see Figure 2 of Aharonian et al. 2007a). Contours are plotted every 10 smoothed counts from 20 smoothed counts. (b) Averaged brightness temperature distribution of $^{12}$CO($J$=1--0) emission in a velocity range of $V_{\mathrm{LSR}}$ = $-20$ km s$^{-1}$ to $0$ km s$^{-1}$ is shown in color (Fukui et al. 2003, Moriguchi et al. 2005). $White$ $contours$ show the H.E.S.S. TeV $\gamma$-ray distribution and are plotted every 20 smoothed counts from 20 smoothed counts. (c) Averaged brightness temperature distribution of H{\sc i} emission obtained by ATCA and Parkes in a limited velocity range from $V_{\mathrm{LSR}}$ = $-8$ km s$^{-1}$ to $-6$ km s$^{-1}$ (McClure-Griffiths et al. 2005) is shown in color. $White$ $contours$ show the $^{12}$CO($J$=1--0) brightness temperature integrated in the same velocity range every 1.0 K km s$^{-1}$ ($\sim$3$\sigma$) (Fukui et al. 2012).}
\end{center}
\end{figure*}%

\begin{figure*}[t]
\begin{center}
\includegraphics[width=117mm,clip]{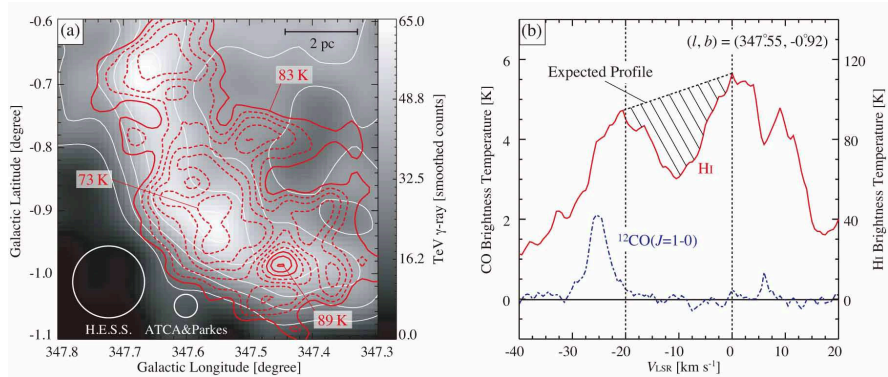}
\caption{(a)The H.E.S.S. TeV $\gamma$-ray distribution toward the SE cloud (Aharonian et al. 2007a). $Red$ $contours$ show averaged H{\sc i} brightness temperature distribution in a velocity range from $-$15 km s$^{-1}$ to $-$5 km s$^{-1}$ (McClure-Griffiths et al. 2005). (b) The H{\sc i} and $^{12}$CO($J$=1--0) spectra at ($l$, $b$) =(347.55 degrees, $-$0.92 degrees). The shaded area shows an expected H{\sc i} profile. (Fukui et al. 2012)}
\end{center}
\end{figure*}%

Figures 3a-c shows the distributions of TeV $\gamma$-rays, CO and H{\sc i}. The velocity range of the interacting gas with the SNR is later confirmed to be $-20$ km s$^{-1}$ -- 0 km s$^{-1}$ and is shown in Figure 3b (Fukui et al. 2012). We shall adopt this velocity range hereafter. The TeV $\gamma$-rays show a well-defined SNR shell. The CO distribution shows a relatively good correspondence with the TeV $\gamma$-rays, whereas in some places the correspondence breaks. The most notable break is seen toward the southeastern rim (SE rim) of the $\gamma$-ray shell where no CO is seen. The H{\sc i} distribution shows that H{\sc i} is distributed over the entire SNR with a peak in the north, suggesting that a significant amount of H{\sc i} gas is also associated with the SNR. The H{\sc i} intensity shows decrease toward the west and southeast in the SNR. The decrease in the west coincides with the CO and represents low-abundance cold H{\sc i} in the molecular cloud. The decrease in the southeast is different from that, because there is no CO in the SE rim. 

Figure 4a shows the H{\sc i} distribution overlayed on the TeV $\gamma$-ray distribution toward the SE rim and Figure 4b shows a typical H{\sc i} profile and a CO profile in the SE rim at ($l$, $b$) = (347.55 degrees, $-$0.92 degrees). The H{\sc i} contours show a fairly good correlation with the $\gamma$-rays, whereas the H{\sc i} intensity actually decreases toward the $\gamma$-ray enhancements. The H{\sc i} profile shows a broad dip in a velocity range from $-$20 to 0 km s$^{-1}$. We interpret that the dip is due to self-absorption by cool H{\sc i} associated with the SNR. Such a broad dip is unusual for cold H{\sc i} dips in typical dark clouds which have generally a narrow linewidth of a few km s$^{-1}$. In the present case, we ascribe the broad dip as due to the large expanding motion of the H{\sc i} cavity wall. This H{\sc i} gas is not seen in CO, indicating that gas density is lower than 10$^3$ cm$^{-3}$ and that the gas still contains fairly dense H{\sc i} in the order of 100 cm$^{-3}$ as estimated below. The spin temperature $T_{\mathrm{s}}$ of cool H{\sc i} decreases with density due to the increased atomic line cooling and photon shielding as shown by numerical calculations (e.g., Figure 2 in Goldsmith, Li $\&$ Krco 2007). This is just consistent with the behavior in Figure 4a; the decease of the H{\sc i} brightness indicates higher H{\sc i} density in the SE rim and the $\gamma$-ray count increases with the density increase. By taking reasonable $T_\mathrm{s}$ ($\sim$40 K) as inferred from the minimum intensity of H{\sc i} at the dip bottom, we calculate the H{\sc i} column density responsible for the self-absorption. Here we adopt a conventional assumption that the background H{\sc i} emission is interpolated by a straight line (Figure 4b). We estimate that the H{\sc i} optical depth is close to 1 and calculate the absorbing H{\sc i} column density to be $\sim$3$\times$10$^{21}$ cm$^{-2}$ (see Figure 5b), indicating that the SE rim has enhanced H{\sc i} density. As shown later in Figure 7, this interpretation of the H{\sc i} dip is consistent with the visual extinction which also shows enhanced ISM toward the SE rim.

\begin{figure*}[t]
\begin{center}
\includegraphics[width=117mm,clip]{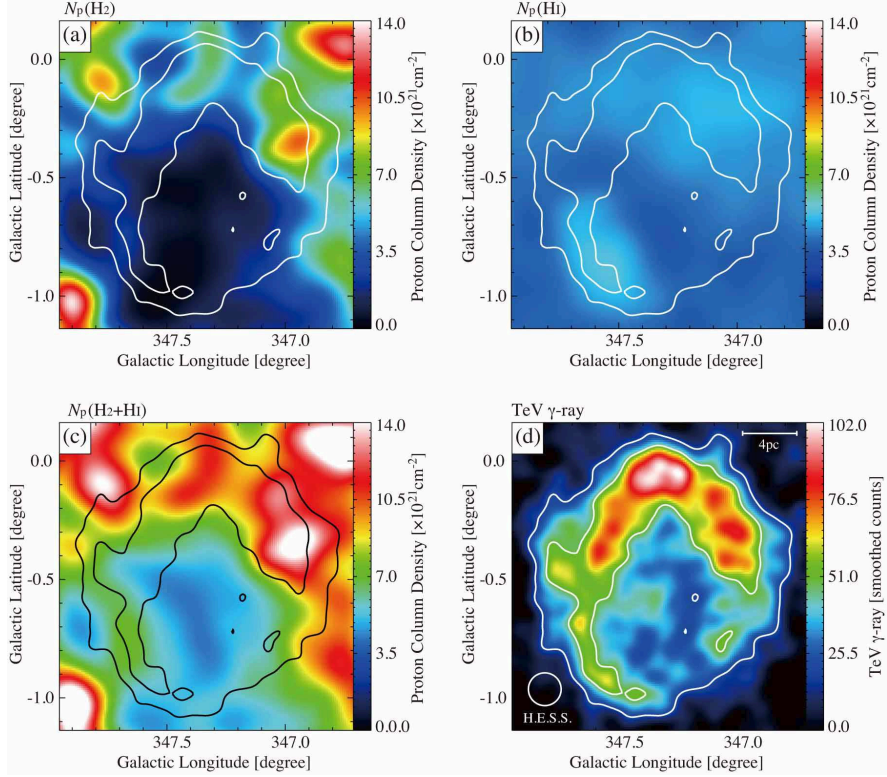}
\caption{(a) Distributions of column density of ISM protons $N_{\mathrm{p}}$ estimated from $^{12}$CO($J$=1--0), $N_{\mathrm{p}}$(H$_2$), (b) H{\sc i} emission with correction for the H{\sc i} self-absorption $N_{\mathrm{p}}$(H{\sc i}) and (c) $N_{\mathrm{p}}$(H$_2$+H{\sc i}). All the datasets used here are smoothed to FWHM of TeV $\gamma$-ray PSF. (d) TeV $\gamma$-ray distribution. Contours are plotted every 50 smoothed counts from 20 smoothed counts (Fukui et al. 2012).}
\end{center}
\end{figure*}%

We present below some additional details on derivation of the ISM proton distribution. The molecular column density is calculated by using an $X$ factor to convert the $^{12}$CO $J$=1--0 intensity into H$_{\mathrm{2}}$ column density (equation 1) and the error is due to the 3$\sigma$ noise fluctuations in the CO observations. 

\begin{equation}
N(\mathrm{H_2}) = X \cdot W(^{12}\mathrm{(CO)})
\end{equation}

where the $X$ factor, $N$($\mathrm{H_2}$) / $W$($^{12}$CO)(K km s$^{-1}$), adopted is 2.0 $\times$ 10$^{20}$ [cm$^{-2}$  / (K km s$^{-1}$)] (Bertsch et al. 1993). Then, the molecular proton density is given as $N_\mathrm{p}$ ($\mathrm{H_2})$ = 2$N$($\mathrm{H_2})$.

The peak H{\sc i} brightness is usually around 80--100 K in the region except for the regions with H{\sc i} self-absorption. The brightness is consistent with warm H{\sc i} with typical spin temperature $T_{\mathrm{s}}$ $\sim$125 K. The atomic proton column density is estimated by assuming that the 21 cm H{\sc i} line is optically thin as follows (Dickey $\&$ Lockman 1990);
\begin{equation}
N_{\mathrm{p}}{(\mbox{H{\sc i}})} = 1.823 \times 10^{18} \int T_\mathrm{b} dV \:\: \mathrm{(cm^{-2})}
\end{equation}

where $T_\mathrm{b}$ (K) and $V$ (km s$^{-1}$) are the H{\sc i} brightness and velocity. The self-absorbing H{\sc i} column density is added to this in order to obtain the total atomic protons. The noise fluctuations in the H{\sc i} brightness are very small. The major uncertainty here comes from the uncertainty of $T_\mathrm{s}$ of the cool H{\sc i} and the interpolation of the background H{\sc i}, but it is smaller than 10$^{21}$ cm$^{-2}$ and do not significantly affect the present estimates of $N_{\mathrm{p}}$(H$_2$+H{\sc i}).

The total ISM proton column density is finally given by the sum as follows; 
\begin{equation}
N_{\mathrm{p}}(\mathrm{H_2}+\mbox{H{\sc i}}) = N_{\mathrm{p}}(\mathrm{H_2}) + N_{\mathrm{p}}(\mbox{H{\sc i}})
\end{equation}The average $\mathrm{H_2}$ density of the CO cloud is around several 100 cm$^{-3}$ and that of the SE rim is around 100 cm$^{-3}$. 

\begin{figure*}[t]
\begin{center}
\includegraphics[width=117mm,clip]{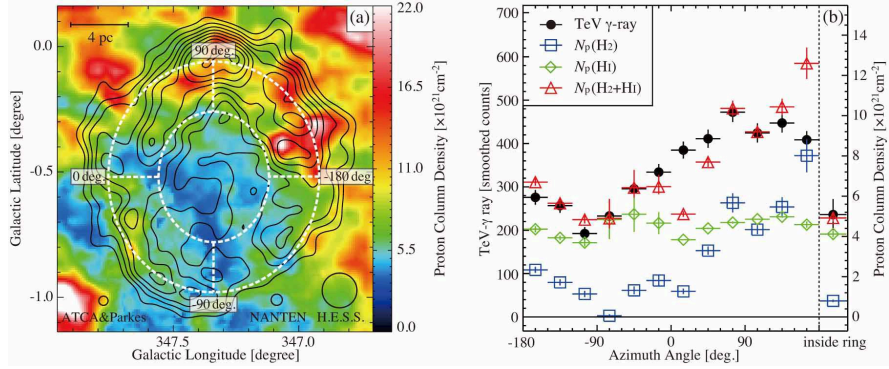}
\caption{(a) Distributions of column density of the total ISM protons $N_{\mathrm{p}}$(H$_2$+H{\sc i}) in a velocity range from $-$20 km s$^{-1}$ to 0 km s$^{-1}$. Contours are the same as in Figure 1(a). (b) Azimuthal distributions of $N_{\mathrm{p}}$(H$_2$), $N_{\mathrm{p}}$(H{\sc i}), $N_{\mathrm{p}}$(H$_2$+H{\sc i}) and TeV $\gamma$-ray smoothed counts per beam between the two elliptical rings shown in Figure 8(a). The proton column densities are averaged values between the rings (see text). Semi-major and semi-minor radii of the outer ring are 0.46 degrees and 0.42 degrees, respectively, and the radii of the inner ring are half of them. The same plots inside the inner ring are shown on the right side of Figure 8(b) (Fukui et al. 2012).}
\end{center}
\end{figure*}%

Figure 5 summarizes the results. Figures 5a--d shows $N_{\mathrm{p}}$(H$_2$), $N_{\mathrm{p}}$(H{\sc i}), $N_{\mathrm{p}}$(H$_2$+H{\sc i}) and TeV $\gamma$-rays, respectively. First, we see that there is a significant amount of $N_{\mathrm{p}}$(H{\sc i}) (column density is several times 10$^{21}$ cm$^{-2}$) over the SNR particularly toward the SE rim. The total ISM protons shows a significant difference by inclusion of the H{\sc i} and the ISM proton distribution shows a much better correspondence with the $\gamma$-ray distribution than the H$_2$ alone. Figure 6 shows a comparison of the azimuthal angular distribution between $N_{\mathrm{p}}$ and $\gamma$-rays and we see $N_{\mathrm{p}}$(H$_2$+H{\sc i}) shows a good correspondence with the TeV $\gamma$-rays. This is the first result which demonstrates a good spatial correspondence between TeV $\gamma$-rays and ISM protons at 0.1 degree angular resolution and proves that a necessary condition for the hadronic scenario is fulfilled in the TeV $\gamma$-ray SNR. One of the natural next steps is to extend the method to the other SNRs as shown for RX J0852.0$-$4622 in the next sub-section.

\begin{figure*}[t]
\begin{center}
\includegraphics[width=117mm,clip]{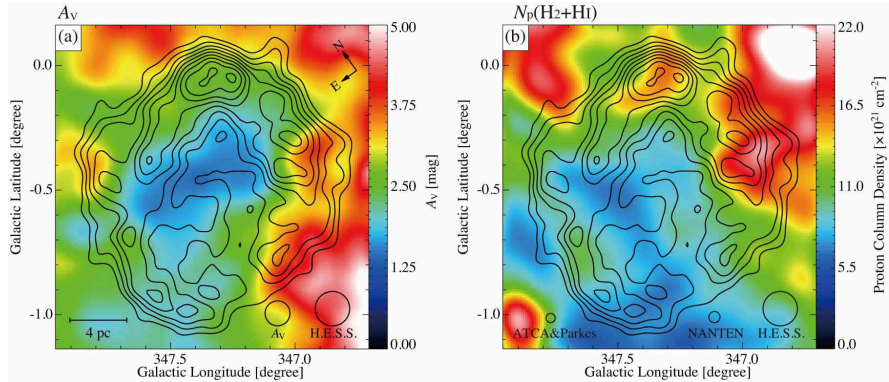}
\caption{(a)$A_{\mathrm{V}}$ distribution (Dobashi et al. 2005) is shown in color. TeV $\gamma$-ray contours are the same as in Figure 1(a). (b) Distribution of column density of the total ISM protons estimated from both CO and H{\sc i} in a velocity range from $V_{\mathrm{LSR}}$ = $-20$ to 10 km s$^{-1}$. Here the H{\sc i} self-absorption is taken into account. TeV $\gamma$-ray contours are the same as in Figure 1(a) (Fukui et al. 2012).}
\end{center}
\end{figure*}%

An important implication of the H{\sc i} analysis is that the SE rim has enhanced column density in H{\sc i} as in the TeV $\gamma$-rays, whereas the analysis of the H{\sc i} dip has some uncertainty in $T_{\mathrm{s}}$ and the H{\sc i} background interpolation. In order to make an independent check of the H{\sc i} column density, we can use other methods to measure the ISM protons. The visual extinction is one of such methods applicable at a distance of 1 kpc. Figure 7a shows that the distribution of the visual extinction is consistent with the H{\sc i} result in the SE rim, lending another support for the analysis of the H{\sc i} dip as self-absorption. Figure 7b shows for comparison the sum of the ISM proton distribution including both the foreground ISM protons and those in the SNR. More recently, a new analysis of the Suzaku X-ray data also shows a similar enhanced absorption column density toward the SE rim (Sano et al. 2013b). Considering all these, we are confident that the $N_{\mathrm{p}}$(H$_2$+H{\sc i}) distribution in Figure 6a is reliable.

\subsection{RX J0852.0$-$4622}
\label{subsec:3-3}
RX J0852.0$-$4622 is another prominent young TeV $\gamma$-ray SNR imaged by H.E.S.S. Aschenbach et al. (1998) discovered RX J0852.0$-$4622 as a hard X-ray SNR seen in projection against the Vela SNR in ROSAT All-Sky Survey image. RX J0852.0$-$4622 and RX J1713.7$-$3946 share similar properties; they are both young with ages of 1600--2000 yrs and show non-thermal synchrotron shell-like X-ray emission with no thermal features in addition to the strong shell-like TeV $\gamma$-rays (Figure 8a). The apparent large diameter of RX J0852.0$-$4622, about 2 degrees, is most favorable to test a spatial correspondence between the $\gamma$-rays and the ISM at a 0.12 degrees angular resolution (FWHM) of H.E.S.S.. There are two more similar shell-like TeV $\gamma$-ray SNRs, RCW 86 and HESS J1731$-$347, but they are small with a size of about 0.5 degrees (Aharonian et al. 2009; H.E.S.S.Collaboration et al. 2011), making it difficult to test the spatial correspondence. 

The age and distance of RX J0852.0$-$4622 have been the subject of debate in the literature. Slane et al. (2001a, b) argued that RX J0852.0$-$4622 is physically associated with the GMC, the Vela Molecular Ridge (VMR, May et al. 1988; Yamaguchi et al. 1999). The distance of the VMR is estimated to be 700 $\pm$ 200 pc (Liseau et al. 1992), whereas it is not yet established if the VMR is physically connected to RX J0852.0$-$4622 (see e.g., Pannuti et al. 2010).  Recently, Katsuda et al. (2008) estimated an expansion rate of RX J0852.0$-$4622 based on two observations separated by 6.5 years made with XMM-Newton toward the northwestern rim of RX J0852.0$-$4622, and derived an age of 1.7--4.3 $\times$ 10$^3$ years for the SNR and a distance of 750 pc. We will adopt this distance to RX J0852.0$-$4622. 

\begin{figure*}[t]
\begin{center}
\includegraphics[width=117mm,clip]{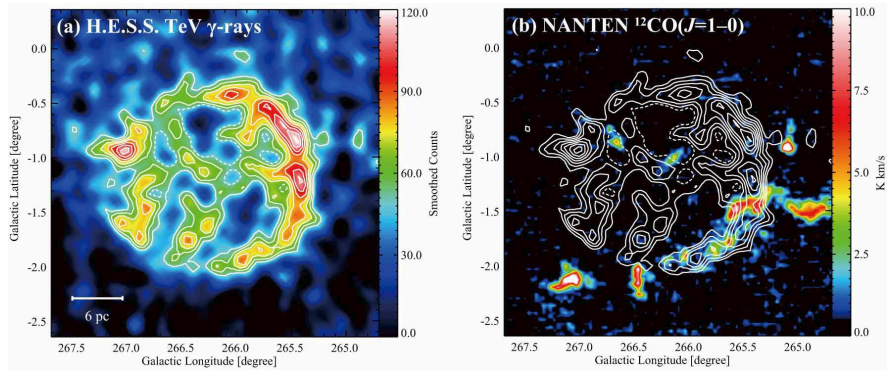}
\caption{(a) The H.E.S.S. TeV $\gamma$-ray distribution of RX J0852.0$-$4622 in smoothed excess counts (see Figure 1 of Aharonian et al. 2007b). Contours are plotted every 10 smoothed counts from 50 smoothed counts. (b) Integrated intensity distribution of $^{12}$CO($J$=1--0) emission in a velocity range of $V_{\mathrm{LSR}}$ = 24 km s$^{-1}$ to 30 km s$^{-1}$ is shown in color (Moriguchi et al. 2001). $White$ $contours$ show the H.E.S.S. TeV $\gamma$-ray distribution and are plotted the same as Figure 8 (a).}
\end{center}
\end{figure*}%

In the region of RX J0852.0$-$4622, the CO clouds are significantly less than in the region of RX J1713.7$-$3946. This is not surprising since RX J0852.0$-$4622 is far from the Galactic center where the ISM density is generally lower than in the region of RX J1713.7$-$3946. Among all the CO features in the direction, we find that the CO clouds at 25 km s$^{-1}$ show clear association with the southwestern rim of the SNR (Figure 8b). Based on this correspondence we have searched for associated CO and H{\sc i} gas with the SNR shell. Figure 9a shows the total ISM proton distribution in RX J0852.0$-$4622 derived by combining the associated CO and H{\sc i} distributions. The velocity range is from 0 to 50 km s$^{-1}$ centered at 25 km s$^{-1}$, although the Galactic rotation suggests lower velocity less than 10 km s$^{-1}$. We ascribe this velocity deviation as due to the expansion of a few H{\sc i} supershells; the SNR RX J0852.0$-$4622 is located toward an overlapping region among these H{\sc i} shells and the associated ISM has significant deviation by 50 km s$^{-1}$ at maximum from the pure Galactic rotation. The total associated H{\sc i} mass is 10$^4$ $M_{\odot}$ while that of molecular gas is 10$^3$ $M_{\odot}$. So, the associated ISM is dominated by the atomic gas in RX J0852.0$-$4622. On the other hand, the associated ISM mass in RX J1713.7$-$3946 is 10$^4$ $M_{\odot}$ both in H$_2$ and H{\sc i}, respectively. RX J1713.7$-$3946 is therefore associated with 10 times more molecular gas than RX J0852.0$-$4622.

We made a Gaussian fitting to the radial TeV $\gamma$-ray distribution and determined the best-fit parameters like the radius and width of the shell (Figure 9a). Figure 9b shows the azimuthal TeV $\gamma$-ray distribution (Fukui et al. 2013). This is obtained by averaging the $\gamma$-ray count between the two rings at the 1/3 level of the peak of the Gaussian shell, 0.61 degrees and 0.92 degrees in the radial distribution from the SNR center. The two rings include the major part of the $\gamma$-ray shell and Figure 9b represents well the $\gamma$-ray azimuthal distribution. We find the correspondence between the total ISM and the TeV $\gamma$-rays is reasonably good. This is the second case that shows a good correspondence between the ISM protons and TeV $\gamma$-rays. Toward the angle 15 degrees in Figure 9b there is some excess in the $\gamma$-rays. This is due to a PWN located by chance toward this part of the shell, which and is not associated with the SNR.

\begin{figure*}[t]
\begin{center}
\includegraphics[width=117mm,clip]{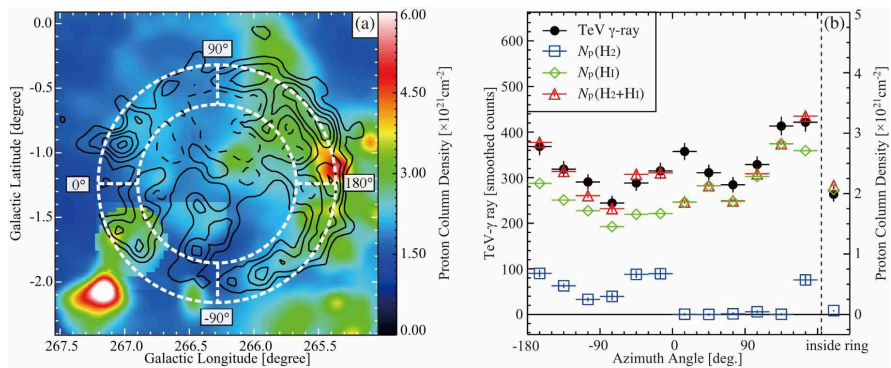}
\caption{(a) The column density of the ISM proton distribution $N_{\mathrm{p}}$($\mathrm{H_2}$+H{\sc i}) and the contours of TeV $\gamma$-rays. The $outer$ and $inner$ $circles$ are centered at ($l$, $b$) = (266.28 degrees, $-1.24$ degrees) with radii of 0.92 degrees and 0.63 degrees, respectively. (b) The azimuthal angles are indicated. The azimuthal distribution of the TeV $\gamma$-rays and the ISM protons. The $\gamma$-ray counts and the column density of the ISM protons in the two circles in Figure 9 (a) are shown at every 30 degrees. The $black$ $dots$ show the TeV $\gamma$-ray counts and the $blue$ $box$ the molecular protons $N_{\mathrm{p}}(\mathrm{H_2})$, the $green$ $box$ the atomic protons $N_{\mathrm{p}}$(H{\sc i}), and the $red$ $triangle$ the total ISM protons $N_{\mathrm{p}}$($\mathrm{H_2}$+H{\sc i}). The right-most point shows the average within the $inner circle$ (Fukui et al. 2013).}
\end{center}
\end{figure*}%

To summarize this section, we have shown that the ISM proton distribution $N_{\mathrm{p}}$($\mathrm{H_2}$+H{\sc i}) shows a good spatial correspondence with the TeV $\gamma$-ray distribution in the two outstanding TeV $\gamma$-ray SNRs. This is a step forward in establishing the hadronic origin of $\gamma$-rays, whereas the results alone do not reject the leptonic scenario. What remains to be explored is 1) if a leptonic model for the $\gamma$-ray production can also explain the observed TeV $\gamma$-ray distribution, and 2) if such a hadronic model is consistent with the non-thermal X-ray distribution by the CR electrons.

\section{Interaction between the ISM and the SNR shock, and the $\gamma$-ray production}
\label{sec:4}

\begin{figure*}[b]
\begin{center}
\includegraphics[width=117mm,clip]{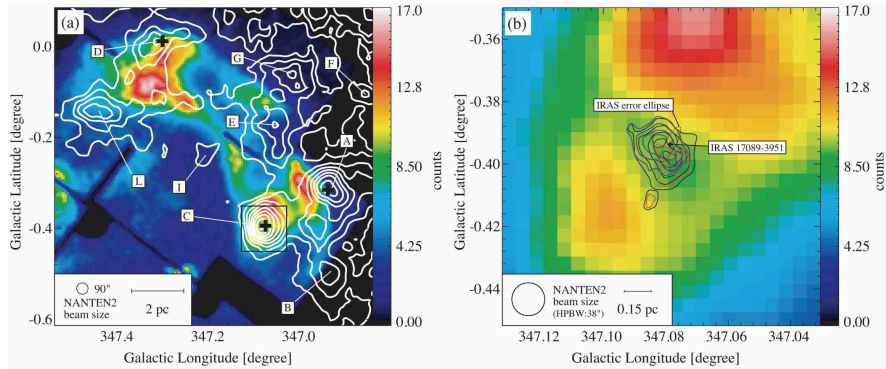}
\caption{$Suzaku$ XIS (XIS 0+2+3) mosaic image of RX J1713.7$-$3946 in the energy bands 1--5 keV in color scale and (a) $^{12}$CO($J$=2-1) (b) $^{12}$CO($J$=4--3) intensity contours obtained by NANTEN2 (Sano et al. 2010 Figure 5). The area enclosed by the $black box$ in (a) is shown enlarged in Figure 9 (b). Open crosses show the positions of the $IRAS$ point sources.}
\end{center}
\end{figure*}%

The comparisons between $\gamma$-rays and the ISM in the two SNRs in Section 3 have led to a picture that the shell-like clumpy ISM consisting of atomic and molecular gas acts as the target for CR protons to produce the hadronic $\gamma$-rays. We shall here discuss into more detail the interaction between the SN shock front and the ISM in RX J1713.7$-$3946. We presume that a similar argument is applicable to RX J0852.0$-$4622 by considering its properties similar to RX J1713.7$-$3946 (Section 3).

Figure 10a shows a comparison between the non-thermal X-rays and the CO distribution in the northwestern (NW) rim of RX J1713.7$-$3946 (Sano et al. 2010). In the X-ray distribution, the NW rim is particularly intense and this is where the densest CO clumps are located.  Figure 10b shows a close-up view toward peak C where a protostar with bipolar outflow is located in the center. Figure 10 shows that the X-rays are correlated with CO at pc scale but is anti-correlated at sub-pc scale as shown by that peak C is located toward the local minimum of the X-rays. This behavior is seen in every CO peak within this SNR (Sano et al. 2013a), offering a hint on the details of the interaction. 

The initial ISM distribution prior to the SN explosion consists of a wind-evacuated low-density cavity of $\sim$9 pc radius and the cavity wall consists of around ten dense molecular clumps, the cool and dense H{\sc i} gas as in the SE rim, and the warm H{\sc i} (Figures 10 and 11). The interaction with the SN shock waves is taking place in the last several 100 yrs as estimated from a ratio of $\sim$3 pc / 3000 km s$^{-1}$. In such a short time scale it is impossible that the molecular clumps are pushed appreciably in space and it is likely that the neutral gas distribution, in particular the molecular distribution, is not significantly affected by the shock interaction. Some of the dense CO clumps located in the inner part of the shell including peak C are likely overtaken by the shock (Sano et al. 2010). In addition, the stellar winds likely have stripped off the lower density envelope of the CO clumps, making a steep density gradient at the clump surface. Such a gradient is in fact observed toward peak C by using the sub-mm transitions of CO to trace the radial molecular density distribution (Sano et al. 2010).

The SN shock front having 3000 km s$^{-1}$ expansion velocity interacts with the ambient highly clumpy ISM distribution. Many of the previous works on the shock interaction in the SNR assume that the ISM is uniform in density for simplicity (e.g., Ellison et al. 2010). If we assume such uniform ISM, in order to produce the observed $\gamma$-rays the ISM density has to be increased to at least 10 cm$^{-3}$. Then, the uniform model predicts that strong thermal X-rays will be emitted by the heating of the ISM in the shock waves. This creates a serious discrepancy with observations since no thermal X-rays are detected. This was one of the major arguments against the hadronic scenario in the RX J1713.7$-$3946 (e.g., Ellison et al. 2010). While the uniform ISM may be possible as a first approximation, more accurate properties of the interaction strongly depend on the ISM density inhomogeneity. We should therefore adopt more realistic highly inhomogeneous ISM distribution as observed in CO (Figure 3). 

\begin{figure*}[b]
\begin{center}
\includegraphics[width=78mm,clip]{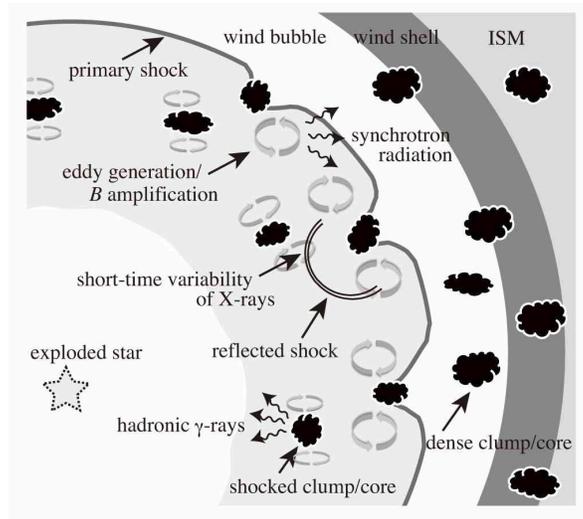}
\caption{Schematic picture of the shock-cloud interaction model (Inoue et al. 2012 Figure 10).  The primary forward shock wave propagates through the cloudy wind bubble, where particle acceleration operates by the diffusive shock acceleration (DSA). The shock waves in the clouds are stalled, which suppresses thermal X-ray line emission. Shock-cloud interactions induce turbulent eddies, which amplifies the magnetic field ($\sim$1 mG) caused by the turbulent dynamo effect and the reverse shocks. Then the synchrotron emissions enhanced around the clouds where magnetic field strength is $\sim$1 mG around shocked clouds. Also, hadronic $\gamma$-rays are emitted from dense clouds illuminated by accelerated protons whose photon index can be $p-$1/2 = 1.5 for $p$ = 2.}
\end{center}
\end{figure*}%

\begin{figure*}
\begin{center}
\includegraphics[width=117mm,clip]{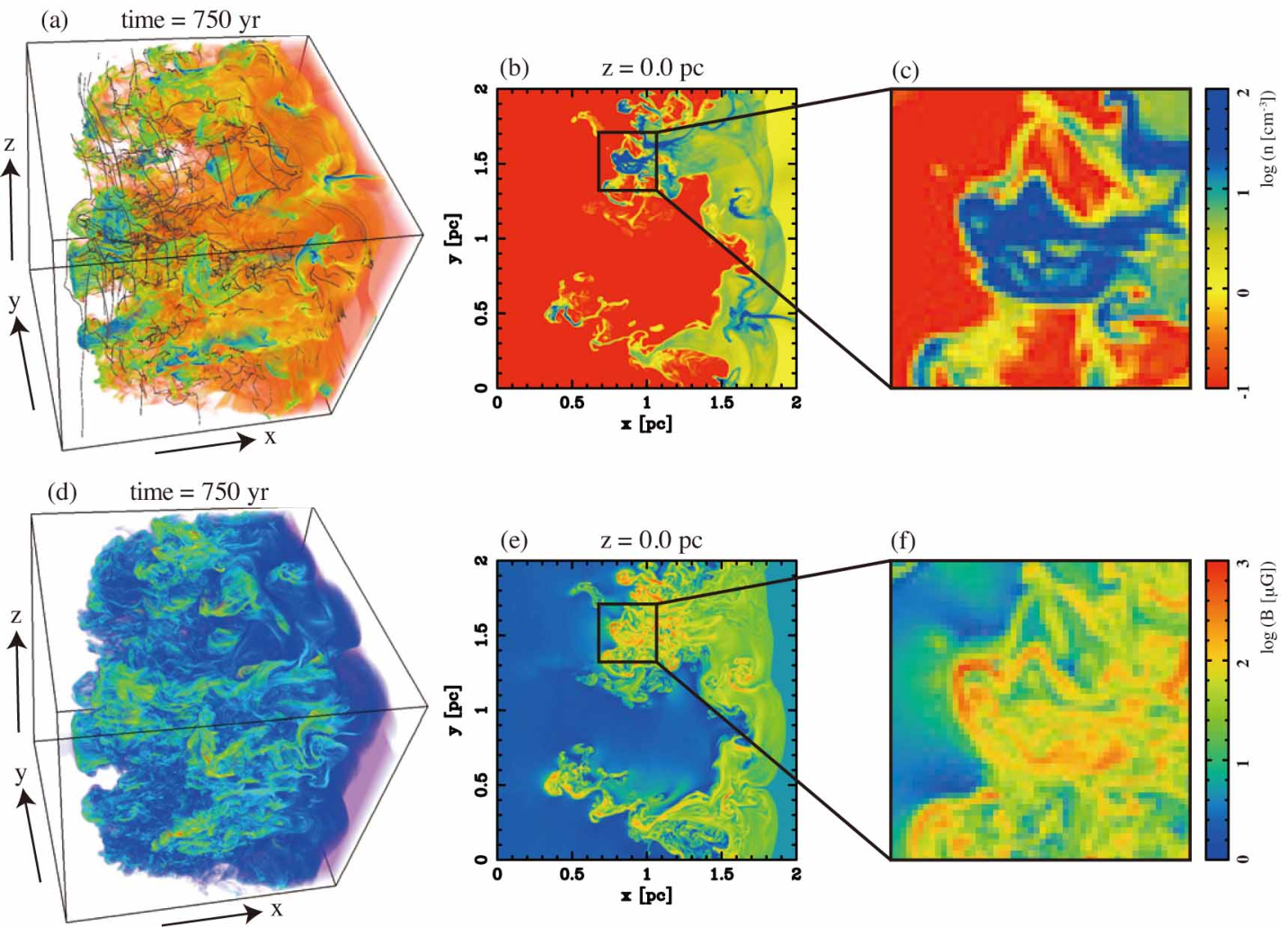}
\caption{Results of the numerical simulations between the SN shockwaves on the ISM (Inoue et al. 2012). The perpendicular shock case at $t$ = 750 yr after the shock injection (Inoue et al. 2012). (a): number density volume rendering. Regions in $green$ and $blue$ indicate the density $n \sim$10 cm$^{-3}$ and $n \geq$ 30 cm$^{-3}$, respectively, and the regions in warm colors show the shocked diffuse gas with $n \leq$ 4 cm$^{-3}$. Magnetic field lines are represented as gray lines. (b): two-dimensional number density slice at $z$ = 0.0 pc. (c): close up of the $solid$ $box$ in (b). (d): volume rendering of magnetic field strength. (e): slice of magnetic field strength at $z$ = 0.0 pc. (f): close up of the $solid$ $box$ in (e).}
\end{center}
\end{figure*}%

Inoue et al. (2012) carried out MHD numerical simulations by considering the clumpy ISM distribution consistent with the observations in RX J1713.7$-$3946 and discussed the basic aspects of the interaction and the $\gamma$-ray and X-ray properties. The interaction between the shock and the CO clumps creates turbulence as simulations indicate (Figure 12). This turbulence amplifies the magnetic field around the clumps and enhances the synchrotron X-rays. The magnetic field is in the order of 10--100 $\mu$G on the average but may be enhanced to the mG order (Uchiyama et al. 2007). The shock waves cannot penetrate into the dense molecular clumps and cannot heat them up. This explains the lack of thermal X-rays which happens usually in the other SNRs of lower-density environments. The CR electrons cannot penetrate into the dense clumps since its penetration depth is only in the order of $\sim$0.1 pc in a given timescale $\sim$1000 yrs (Fukui et al. 2012). As a result, the X-rays are enhanced around the CO clump but are depressed inside the clumps. This offers an explanation of the pc scale correlation and sub-pc scale anti-correlation between CO and X-rays (Figure 10). The inside of the cavity is as a whole of very low density due to the stellar winds evacuation, where particle acceleration is efficiently made via diffusive shock acceleration (DSA) to the energy close to the knee (e.g., Bell 1978; Blandford $\&$ Ostriker 1978 ). CR protons have larger penetration lengths of $\sim$1 pc than electrons and can interact with the ISM protons inside the molecular clumps. Such proton-proton reaction explains the $\gamma$-ray production and is consistent with the spatial correspondence between the $\gamma$-rays and the ISM protons shown in Figures 6 and 9. The inhomogeneous ISM therefore solves the difficulties in the uniform ISM models by allowing the $\gamma$-ray production in the dense clumps as well as the suppression of the thermal X-rays.

In order to have a deeper insight into the $\gamma$-ray production, we need to understand the $\gamma$-ray energy spectrum. It is usually assumed that CR protons can interact with all the target ISM protons in a SNR, but, to be realistic, penetration of CR protons into the molecular gas is energy dependent for a high density regime like $\sim$10$^4$ cm$^{-3}$ or more. The high-energy CR protons can penetrate into the dense clumps, whereas the low-energy protons cannot (Gabici et al. 2007, Zirakashvili \& Aharonian 2010). This energy dependent penetration makes the lower energy protons interact only with the lower density ISM protons. This reflects in the $\gamma$-ray energy spectrum and a harder $\gamma$-ray spectrum is expected than in the uniform ISM case.

\begin{figure*}[b]
\begin{center}
\includegraphics[width=117mm,clip]{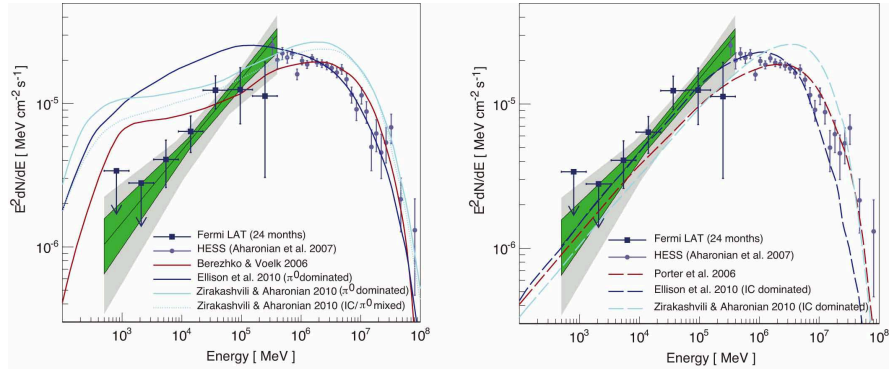}
\caption{Energy spectrum and numerical models of RX J1713.7$-$3946 in GeV to TeV $\gamma$-ray spectra taken by the Fermi Large Area Telescope and H.E.S.S. (Abdo et al. 2011 Figure.3). Some $solid$ $lines$ indicate the numerical model of the hadronic model ($\pi^0$-decay spectrum: $left$) and the leptonic model (inverse Compton scattering spectrum: $right$). The $green$ region shows the uncertainty band obtained from maximum likelihood fit of the spectrum assuming a power law between 500 MeV and 400 GeV for the default model of the region. The $gray$ region depicts the systematic uncertainty of this fit obtained by variation of the background and source models. The $black$ error bars correspond to independent fits of the flux of RX J1713.7$-$3946 in the respective energy bands. Upper limits are set at a 95$\%$ confidence level.}
\end{center}
\end{figure*}%
Figures 13 shows the $\gamma$-ray spectrum of RX J1713.7$-$3946 covering the energy range 10$^3$--10$^8$ MeV (Abdo et al. 2011). This is the combined data of H.E.S.S. and Fermi satellite (Aharonian et al. 2007a; Abdo et al. 2011). RX J1713.7$-$3946 shows a hard spectrum which is not compatible with the conventional hadronic $\gamma$-rays and these authors claimed that the spectrum supports the leptonic scenario. The density- and energy-dependence of the hadronic interaction however indicates that such a hard spectrum is also explained well in the hadronic scenario with highly inhomogeneous ISM distribution as argued into detail by Inoue et al. (2012). Figure 14 shows the $\gamma$-ray spectrum in the same energy band with Figure 13 for RX J0852.0$-$4622 (Tanaka et al. 2011). This shows a softer spectrum than in Figure 13 and may be better explained by the hadronic scenario than the leptonic scenario. In connection with the ISM distribution, we note that RX J0852.0$-$4622 has much less CO clouds than in RX J1713.7$-$3946. CR protons will therefore interact mainly with the H{\sc i} gas whose mass is 10 times more than the molecular mass, offering an explanation of the softer spectrum by the hadronic scenario. These two cases suggest that the ISM distribution is crucial in interpreting the $\gamma$-ray properties and, in particular, in exploring the hadronic $\gamma$-rays.

\begin{figure*}[t]
\begin{center}
\includegraphics[width=117mm,clip]{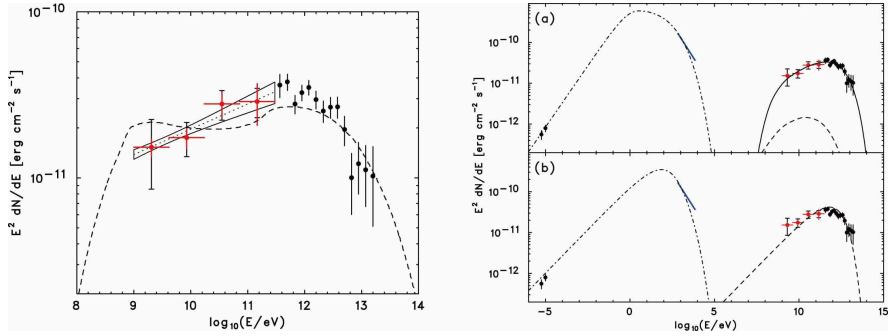}
\caption{Energy spectrum and numerical models of RX J0852.0$-$4622 in GeV to TeV $\gamma$-ray spectra taken by the Fermi Large Area Telescope ($red$) and H.E.S.S. ($black$) (Tanaka et al. 2011 Figure.2 and 3). For the Fermi-LAT points in $left$ $panel$, the $vertical$ $red$ $lines$ and the $black$ $caps$ indicate statistical and systematic errors, respectively. The $dotted$ $line$ shows the best-fit power-law obtained from the maximum likelihood fit for the entire 1--300 GeV band (The butterfly shape is the 68$\%$ confidence region). The $dashed$ $curve$ is the $\pi^0$-decay spectrum (hadronic model) by Berezhko et al. (2009). Broadband SED of RX J0852.0$-$4622 with (a) the hadronic model and (b) the leptonic model. The radio data points and $blue$ $line$ indicate the integrated fluxes of the SNR determined based on the 64 m Parkes radio telescope (Duncan $\&$ Green 2000) and the the X-ray flux using the ASCA data-set (Aharonian et al. 2007b). The $solid$, $dashed$, and $dot$-$dashed$ $lines$ represent contributions from $\pi^0$-decays, inverse Compton scattering, and synchrotron radiation, respectively.}
\end{center}
\end{figure*}%

The identification of target ISM protons in the hadronic scenario allows us to estimate the total CR proton energy $W_{\mathrm{p}}$ in these SNRs. By using the relationships which connect the $\gamma$-ray counts and the ISM density (Aharonian et al. 2006b, 2007a), we estimate that $W_{\mathrm{p}}$ is 10$^{48}$ ergs in the two SNRs. This energy corresponds to 0.1$\%$ of the total kinetic energy of a SNR and seems to be 10 times less than what is required to explain the CR energy budget of the whole Galaxy. There are a few possibilities to reconcile this discrepancy. The CR energy may yet increase in time with the evolution of a SNR and the total CR energy may become more than estimated above. In W44, a middle-aged SNR, the lower limit of $W_{\mathrm{p}}$ is estimated to be 10$^{49}$ erg, 10 times larger than the present value (Giuliani et al. 2011, Yoshiike et al. 2012). It is also to be considered that the ISM volume filling factor in a SNR shell may not be 100$\%$. Particularly, in RX J0852.0$-$4622 the eastern half of the SNR does not have much ISM, whereas we expect CR protons have a similar number density over the SNR. This suggests that some portion of the CR protons may not be detected in the $\gamma$-rays, leading to an under-estimate of $W_{\mathrm{p}}$ by some factor. 

In order to better understand the origin of $\gamma$-rays in SNRs, we need to have a larger sample of SNRs. A direction is to extend an analysis of the ISM to the middle-aged SNRs like W44, and the other to study smaller TeV $\gamma$-ray SNRs like HESS J1731$-$347 (H.E.S.S.Collaboration et al. 2011). The latter should be possible with the future instrument CTA at a higher angular resolution of 1 arcmin with higher sensitivity. This will definitely help us to increase the number of samples significantly and provide a better understanding of the $\gamma$-ray origin.

\section{Dark gas}
\label{sec:5}
An important topic related to the $\gamma$-rays originating from the ISM protons is “dark gas”. The term “dark gas” is used to represent unseen gas either in H{\sc i} or CO but is detectable as $\gamma$-ray excess (Grenier et al. 2005). The first observational study of dark gas was made by using the EGRET data toward the local ISM and showed that there is $\gamma$-ray excess, at galactic latitude 5 degrees $\leq|b|\leq$ 80 degrees, beyond the level which the gas detected by the CO and H{\sc i} transitions is able to explain. Recently, the Planck satellite provided sensitive mm/sub-mm distributions of the same region and showed that 15$\%$ of the mm/sub-mm dust emission is not explained either by the CO or H{\sc i} gas (Planck Collaboration et al. 2011b). This is also called dark gas. In addition, Planck satellite presents similar dark gas components in the Magellanic Clouds (Planck Collaboration et al. 2011a). It is often suggested that dark gas may corresponds to molecular gas with no detectable CO (Planck Collaboration et al. 2011b), while another possibility discussed is opaque H{\sc i} gas whose intensity is saturated by moderately large optical depth.

The results of the $\gamma$-ray SNR in Section 3 suggest that cool H{\sc i} gas having large optical depth around 1 is responsible for producing part of the $\gamma$-rays via the hadronic process. We here discuss that dark gas may correspond to such cool and dense H{\sc i} gas with relatively large optical depth. Figure 15 shows the H{\sc i} intensity as a function of H{\sc i} column density for two spin temperatures $T_{\mathrm{s}}$ = 100 K and 40 K. The assumption here is that the H{\sc i} linewidth is 10 km s$^{-1}$. The plot shows that the H{\sc i} intensity becomes saturated in the H{\sc i} column density range from 10$^{21}$ cm$^{-2}$ to 10$^{22}$ cm$^{-2}$. For $T_{\mathrm{s}}$ = 40 K, the optical depth of H{\sc i} is greater than 1 and the typical H{\sc i} density is around 100 cm$^{-3}$. These H{\sc i} column densities correspond to that of the dark gas observed by Planck and we suggest that the purely atomic dense gas is a viable alternative to CO-free molecular gas to explain the dark gas. 

\begin{figure*}[t]
\begin{center}
\includegraphics[width=78mm,clip]{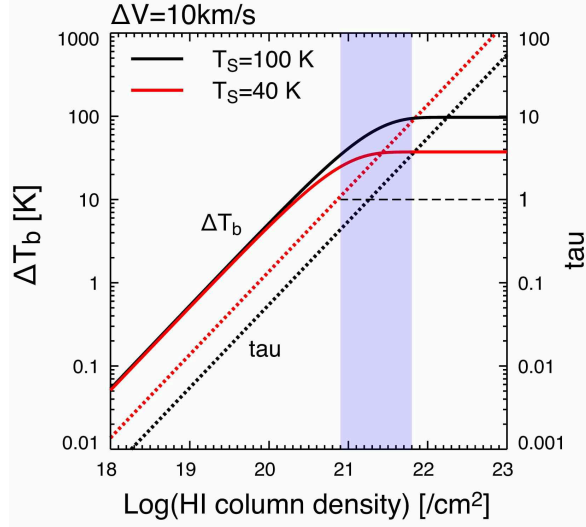}
\caption{$\Delta T_\mathrm{b}$ and $\tau$ variations depending on $N_\mathrm{p}$(H{\sc i}) shown by Torii et al. (2012). Here $\Delta V$ of 10 km s$^{-1}$ is assumed. The filled area shows a $N_\mathrm{p}$ range of the dark gas (Grenier et al. 2005; Planck Collaboration et al. 2011b).}
\end{center}
\end{figure*}%
We discuss on the time scale of H$_2$ formation in order to shed another light on the dark gas composition. The H$_2$ molecules are formed on dust surface with the formation time scale $t_\mathrm{form}$ given by the equation (4) (Hollenbach $\&$ Salpeter 1971).

\begin{equation}
t_\mathrm{form} \sim 10^7 \biggl(\frac{100}{\;n\;\mathrm{(cm^{-3})}} \biggl)\;\mathrm{(yr)}
\end{equation}
For density of $n$ = 100 cm$^{-3}$, the H$_2$ formation timescale is 10 Myrs. This is similar to the typical evolutionary timescale of GMCs. On the other hand, the local gas at high galactic latitude where the dark gas is identified is all low column density clouds whose H{\sc i} column density is in the order of 10$^{21}$ cm$^{-2}$. They are surrounding well known local molecular clouds like Ophiuchus and Lupus. The crossing time scale of such ensembles of local clouds, most likely not gravitationally bound, is estimated to be $\sim$1 Myrs as derived from a ratio 10 pc / 10 km s$^{-1}$ (e.g., see Figure 4 in Tachihara et al., 2001). So, the time scale of the surrounding gas is too short to convert H{\sc i} into H$_2$. This argues for that the dark gas is dominated by dense H{\sc i} but not by H$_2$ without CO. On the other hand, it is conceivable that the envelops of the GMCs in the Galactic plane may have lifetime long enough to form H$_2$, and H$_2$ envelope without CO may be a viable alternative. Such GMCs are not well sampled in the Galaxy due to heavy contamination in the Galactic plane but the GMCs in the LMC (Kawamura et al. 2009) may provide an observable case where the molecular dark gas without CO is dominant. Future careful studies of the dust components will help to clarify the real contents of the dark gas.

\section{Summary}
\label{sec:6}

From a view point of high-energy astrophysics, in particular with an emphasis of the origin of $\gamma$-rays, we present properties of neutral gas in the interstellar medium and its connection with the $\gamma$-ray SNRs. 

We first review the gas properties of both atomic and molecular forms as probed by the 2.6 mm CO and 21 cm H{\sc i} transitions. CO traces dense molecular gas and H{\sc i} traces lower density atomic gas, whereas either CO or H{\sc i} may not be a good tracer in an intermediate density regime roughly between 100 cm$^{-3}$ and 1000 cm$^{-3}$. 

We then present analyses of the two most remarkable TeV $\gamma$-ray SNRs RX J1713.7$-$3946 and RX J0852.0$-$4622 and demonstrate that the ISM proton distribution shows a good spatial correspondence with the TeV $\gamma$-ray distribution. We argue that this correspondence provides a support for the hadronic scenario of $\gamma$-ray production. What remains to be explored is if a leptonic model consistent with the non-thermal X-ray distribution can also explain the observed TeV $\gamma$-ray distribution. We describe theoretical numerical simulations of the interaction of the SNR shock waves with the clumpy ISM and discuss that such realistic simulations can explain the observations reasonably well, whereas the uniform ISM models are not viable.

Then, we present a discussion on the dark gas observed in the $\gamma$-rays and mm/sub-mm dust emission. We argue that, for the local component of the dark gas in the Galaxy, dense H{\sc i} gas as observed in RX J1713.7$-$3946 is a good candidate for the dark gas. The local gas is dominated by low mass clouds and their time scale may be too short to convert H{\sc i} into H$_2$. On the other hand, the GMCs in the LMC have larger time scale like 10 Myrs, long enough for H$_2$ formation, and therefore molecular dominant dark gas is a viable model on the GMC scale.

Finally, we conclude that the ISM plays an essential role in producing the $\gamma$-rays and X-rays in the SNR, and the future instruments like CTA, combined with the ISM studies, will achieve tremendous progress in the $\gamma$-ray astrophysics of the SNR and the other high-energy objects.

\begin{acknowledgement}
The author is grateful to the SOC for inviting him to this most stimulating conference. He is grateful to all the collaborators in this project for their invaluable contributions. His special thanks are to Felix Aharonian, Gavin Rowell, Naomi M. McClure-Griffiths, Tsuyoshi Inoue and Shu-ichiro Inutsuka. NANTEN2 is an international collaboration of 10 universities; Nagoya University, Osaka Prefecture University, University of Cologne, University of Bonn, Seoul National University, University of Chile, University of New South Wales, Macquarie University, University of Sydney, and University of ETH Zurich. This work was financially supported by a grant-in-aid for Scientific Research (KAKENHI, No. 21253003, No. 23403001, No. 22540250, No. 22244014, No. 23740149, No. 22740119 and No. 24224005) from MEXT (the Ministry of Education, Culture, Sports, Science and Technology of Japan). This work was also financially supported by the Young Research Overseas Visits Program for Vitalizing Brain Circulation (R2211) and the Institutional Program for Young Researcher Overseas Visits (R29) by JSPS (Japan Society for the Promotion of Science) and by the grant-in-aid for Nagoya University Global COE Program, “Quest for Fundamental Principles in the Universe: From Particles to the Solar System and the Cosmos,” from MEXT.
\end{acknowledgement}
%

%
%
%

\end{document}